\documentclass[10pt,conference]{IEEEtran}
\IEEEoverridecommandlockouts
\usepackage{cite}
\usepackage{amsmath,amssymb,amsfonts}
\usepackage{algorithmic}
\usepackage{graphicx}
\usepackage{textcomp}
\usepackage{xcolor, colortbl}
\usepackage{booktabs}
\usepackage{multirow}
\usepackage{xspace}
\usepackage{caption}
\usepackage{subcaption}
\usepackage{footnote}
\usepackage{url}
\usepackage{tcolorbox}
\usepackage{booktabs}
\usepackage{enumitem}
\usepackage{fixltx2e}
\usepackage{multicol}
\usepackage{balance}

\usepackage{listings}

\definecolor{Gray}{gray}{0.9}
\definecolor{codegreen}{rgb}{0,0.6,0}
\definecolor{codegray}{rgb}{0.73,0.38,0.06}
\definecolor{codepurple}{rgb}{0.70,0.27,0}
\definecolor{codemagenta}{rgb}{0.74,0.09,0.42}
\definecolor{codeoutput}{rgb}{0.5,0,0}
\definecolor{backcolour}{rgb}{0.96,0.96,0.96}

\def\BibTeX{{\rm B\kern-.05em{\sc i\kern-.025em b}\kern-.08em
    T\kern-.1667em\lower.7ex\hbox{E}\kern-.125emX}}

\usepackage{inconsolata}
\lstdefinestyle{mystyle}{
    backgroundcolor=\color{backcolour},   
    commentstyle=\color{codegreen},
    keywordstyle=\color{codepurple},
    numberstyle=\tiny\color{codegray},
    stringstyle=\color{codemagenta},
    language=Java,
    breakatwhitespace=false,         
    breaklines=true,                 
    keepspaces=true,                 
    numbers=none,                    
    numbersep=5pt,                  
    showspaces=false,                
    showstringspaces=false,
    showtabs=false,                  
    tabsize=2,
    frame=tb,
    framerule=0pt,
    basicstyle=\fontsize{5.5}{5.5}\fontfamily{\ttdefault}\selectfont
}
\lstdefinestyle{mystyleresult}{
    backgroundcolor=\color{backcolour},   
    commentstyle=\color{codegreen},
    keywordstyle=\color{codeoutput},
    numberstyle=\tiny\color{codegray},
    stringstyle=\color{red},
    language=Java,
    breakatwhitespace=false,         
    breaklines=true,                 
    keepspaces=true,                 
    numbers=none,                    
    numbersep=5pt,                  
    showspaces=false,                
    showstringspaces=false,
    showtabs=false,                  
    tabsize=2,
    frame=tb,
    framerule=0pt,
    basicstyle=\color{codeoutput}\fontsize{5.5}{5.5}\fontfamily{\ttdefault}\selectfont
}

\newcommand\revised[1]{\textcolor{black}{#1}}

\begin{document}

\title{\revised{Studying the Usage of Text-To-Text Transfer Transformer to Support Code-Related Tasks}}

\author{
\IEEEauthorblockN{Antonio Mastropaolo\IEEEauthorrefmark{1}, Simone Scalabrino\IEEEauthorrefmark{2}, Nathan Cooper\IEEEauthorrefmark{3}, David Nader Palacio\IEEEauthorrefmark{3}, Denys Poshyvanyk\IEEEauthorrefmark{3},\\Rocco Oliveto\IEEEauthorrefmark{2}, Gabriele Bavota\IEEEauthorrefmark{1}}

\IEEEauthorblockA{\IEEEauthorrefmark{1}\textit{SEART @ Software Institute, Universit\`{a} della Svizzera italiana (USI), Switzerland}}
\IEEEauthorblockA{\IEEEauthorrefmark{2}\textit{University of Molise, Italy}}
\IEEEauthorblockA{\IEEEauthorrefmark{3}\textit{SEMERU @ Computer Science Department, William and Mary, USA}}
}

\maketitle

\newcommand{\ie}{\emph{i.e.,}\xspace}
\newcommand{\eg}{\emph{e.g.,}\xspace}
\newcommand{\etc}{etc.\xspace}
\newcommand{\etal}{\emph{et~al.}\xspace}
\newcommand{\secref}[1]{Section~\ref{#1}\xspace}
\newcommand{\figref}[1]{Fig.~\ref{#1}\xspace}
\newcommand{\listref}[1]{Listing~\ref{#1}\xspace}
\newcommand{\tabref}[1]{Table~\ref{#1}\xspace}
\newcommand{\tool}[1]{{\sc #1}\xspace}

\newcommand{\trainingSentences}{499,618\xspace}
\newcommand{\trainingCode}{1,569,889\xspace}
\newcommand{\bugFixingDataset}{XXX\xspace}
\newcommand{\mutantsDataset}{XXX\xspace}
\newcommand{\assertsDataset}{XXX\xspace}
\newcommand{\commentsDataset}{XXX\xspace}

\newboolean{showcomments}

\setboolean{showcomments}{true}

\ifthenelse{\boolean{showcomments}}
  {\newcommand{\nb}[2]{
    \fbox{\bfseries\sffamily\scriptsize#1}
    {\sf\small$\blacktriangleright$\textit{#2}$\blacktriangleleft$}
   }
  }
  {\newcommand{\nb}[2]{}
  }

\newcommand\TODO[1]{\textcolor{red}{\nb{TODO}{#1}}}

\newcommand\ANTONIO[1]{\textcolor{red}{\nb{ANTONIO}{#1}}}
\newcommand\SIMONE[1]{\textcolor{red}{\nb{SIMONE}{#1}}}
\newcommand\NATHAN[1]{\textcolor{red}{\nb{NATHAN}{#1}}}
\newcommand\DENYS[1]{\textcolor{red}{\nb{DENYS}{#1}}}
\newcommand\ROCCO[1]{\textcolor{red}{\nb{ROCCO}{#1}}}
\newcommand\GABRIELE[1]{\textcolor{red}{\nb{GABRIELE}{#1}}}

\begin{abstract}
Deep learning (DL) techniques are gaining more and more attention in the software engineering community. They have been used to support several code-related tasks, such as automatic bug fixing and code comments generation. \revised{Recent studies in the Natural Language Processing (NLP) field have shown that the Text-To-Text Transfer Transformer (T5) architecture can achieve state-of-the-art performance for a variety of NLP tasks. The basic idea behind T5 is to first pre-train a model on a large and generic dataset using a self-supervised task (\eg filling masked words in sentences). Once the model is pre-trained, it is fine-tuned on smaller and specialized datasets, each one related to a specific task (\eg language translation, sentence classification). In this paper, we empirically investigate how the T5 model performs when pre-trained and fine-tuned to support code-related tasks. We pre-train a T5 model on a dataset composed of natural language English text and source code. Then, we fine-tune such a model by reusing datasets used in four previous works that used DL techniques to: (i) fix bugs, (ii) inject code mutants, (iii) generate assert statements, and (iv) generate code comments. We compared the performance of this single model with the results reported in the four original papers proposing DL-based solutions for those four tasks. We show that our T5 model, exploiting additional data for the self-supervised pre-training phase, can achieve performance improvements over the four baselines.}
\end{abstract}


\section{Introduction} \label{sec:intro}
Deep Learning (DL) has been used to support a vast variety of code-related tasks. Some examples include automatic bug fixing \cite{Tufano:tosem2019,Chen:2019,Mesbah:fse2019,Hata:2018}, learning generic code changes \cite{Tufano:icse2019}, code migration \cite{Nguyen:icse2014,Nguyen:fse2013}, code summarization \cite{LeClair:icse2019,Jiang:ASE'17,Liu:ase2018,haque:2020}, pseudo-code generation \cite{Oda:ase2015}, code deobfuscation \cite{Vasilescu:fse2017,Jaffe:icpc2018}, injection of code mutants \cite{Tufano:icsme2019}, automatic generation of assert statements \cite{Watson:icse2020}, and code completion \cite{Karampatsis:DLareBest,alon2019structural,kim2020code,svyatkovskiy2020intellicode,brody2020neural}. These works customize DL models proposed in the Natural Language Processing (NLP) field to support the previously listed tasks. For instance, Tufano \etal \cite{Tufano:tosem2019} used an RNN Encoder-Decoder architecture, commonly adopted in Neural Machine Translation (NMT) \cite{Kalchbrenner:2013,Sutskever:2014,Cho:2014}, to learn how to automatically fix bugs in Java methods. The model learned bug-fixing patterns by being trained on pairs of buggy and fixed methods mined from software repositories. This work, as the vast majority of the ones previously mentioned (\eg \cite{Jiang:ASE'17,Tufano:icse2019,Tufano:icsme2019,Watson:icse2020,haque:2020}), share one common characteristic: \emph{They shape the problem at hand as a text-to-text transformation, in which the input and the output of the model are text strings}. 

For example, in the work by Watson \etal \cite{Watson:icse2020} the input is a string representing a test method without an assert statement, and the output is an appropriate assert statement for the given test. In the approach by Haque \etal \cite{haque:2020}, the input is composed of strings representing a subroutine to document, while the output is a natural language summary documenting the subroutine. 

Recent years have seen the raise of transfer learning in the field of natural language processing. The basic idea is to first pre-train a model on a large and generic dataset by using a self-supervised task, \eg masking tokens in strings and asking the model to guess the masked tokens. Then, the trained model is fine-tuned on smaller and specialized datasets, each one aimed at supporting a specific task. In this context, Raffel \etal \cite{raffel2019exploring} proposed the T5 (Text-To-Text Transfer Transformer) model, pre-trained on a large natural language corpus and fine-tuned to achieve state-of-the-art performance on many tasks, all characterized by text-to-text transformations.

\revised{The goal of this work is to empirically investigate the potential of a T5 model when pre-trained and fine-tuned to support many of the previously listed code-related tasks also characterized by text-to-text transformations.} We started by pre-training a T5 model using a large dataset consisting of  \trainingSentences English sentences and \trainingCode source code components (\ie methods). 
Then, we fine-tune the model using four datasets from previous work with the goal of supporting four code-related tasks:\smallskip

\emph{Automatic bug-fixing.} We use the dataset by Tufano \etal \cite{Tufano:tosem2019}, composed of instances in which the ``input string'' is represented by a buggy Java method and the ``output string'' is the fixed version of the same method.

\emph{Injection of code mutants.} This dataset is also by Tufano \etal \cite{Tufano:icsme2019}, and features instances in which the input-output strings are reversed as compared to automatic bug-fixing (\ie the input is a fixed method, while the output is its buggy version). The model must learn how to inject bugs (mutants) in code instead of fixing bugs.

\emph{Generation of assert statements in test methods.} We use the dataset by Watson \etal \cite{Watson:icse2020}, composed of instances in which the input string is a representation of a test method without an assert statement and a focal method it tests (\ie the main production method tested), while the output string encodes an appropriate assert statement for the input test method.

\emph{Code Summarization.} We use the dataset by Haque \etal \cite{haque:2020} where input strings are some representations of a Java method to summarize, \& an output string is a textual summary.

Once the T5 model has been fine-tuned on all these tasks, we run it on the same test sets used in the four referenced works \cite{Tufano:tosem2019,Watson:icse2020,haque:2020,Tufano:icsme2019} comparing the achieved results to those reported in the original work. Our results show that the T5 model is able to improve the performance of the original models in all four tasks. 

\revised{Worth noticing is that, besides the different architecture of the T5 model, the latter can take advantage of a pre-training phase in which additional training data is provided as input as compared to the four baselines. This could explain, at least partially, the boost of performance that we observed. Also, as previously said, the additional pre-training is done in a self-supervised way (\ie by simply masking random tokens in the code/text used for pre-training), making this step relatively cheap to perform and scalable to large code bases that can be easily collected from sources such as GitHub. In contrast, the four baselines exploit a completely supervised training (\eg in the case of automatic bug-fixing, the baseline needs pairs of buggy and fixed methods to be trained). Building such a dataset for supervised training has a cost, and there are limitations in terms of the amount of data one can mine.}

\revised{Besides the good performance ensured by the T5, having a single model able to support different tasks can benefit technological transfer since it simplifies the implementation and the maintenance of a tool supporting several tasks.} The code and data used in this work are publicly available \cite{replication}.



 
\section{Related Work} \label{sec:related}
DL techniques have been used to support many software engineering tasks. Due to space limitations, we discuss only the approaches related to the four tasks we subject to our study, with particular attention on those used as baselines. We also introduce notions needed to understand our experimental design. 

\subsection{Automatic Bug-Fixing}
Many techniques have been proposed for the automatic fixing of software bugs. Several of them \cite{LeGoues:tse2012,LeGoues:icse2012,Sidiroglou-Douskos:2015,Pierret:2009,Gabel:2010,Carzaniga:2013,Nguyen:2013:ASE,White:2019:SANER,Bader:oopsla2019} rely on the \emph{redundancy assumption}, claiming that large programs contain the seeds of their own repair. Such an assumption has been verified by at least two independent studies~\cite{Martinez:2014,Barr:2014}. In this section we focus on techniques exploiting DL for bug-fixing. 

Mesbah \etal \cite{Mesbah:fse2019} focus on build-time compilation failures by presenting DeepDelta, an approach using NMT to fix the build. The input is represented by features characterizing the compilation failure (\eg kind of error, AST path, \etc). As output, DeepDelta provides the AST changes needed to fix the error. In the presented empirical evaluation, DeepDelta correctly fixed 19,314 out of 38,788 (50\%) compilation errors.

Chen \etal \cite{Chen:2019} present SequenceR, a sequence-to-sequence approach trained on over 35k single-line bug-fixes. SequenceR takes as input the buggy line together with its ``abstract buggy context'', meaning the relevant code lines from the buggy class. 

The output of the approach is the recommended fix for the buggy line. The approach, tested on a set of 4,711 bugs, was able to automatically fix 950 ($\sim$20\%) of them. Similar approaches have been proposed by Hata \etal \cite{Hata:2018} and Tufano \etal \cite{Tufano:tosem2019}. The latter is the one we compared our approach with and, thus, we describe it in more details.

Tufano \etal \cite{Tufano:tosem2019} investigate the performance of an NMT-based approach in the context of automatic bug-fixing. 

They train an encoder-decoder model on a set of bug-fix pairs (BFPs), meaning pairs of strings in which the first one (input) represents a Java method that has been subject to a bug-fixing activity, and the second one (target) represents the same Java method once the bug was fixed. 

\begin{figure}[h]
	\centering
	\includegraphics[width=0.65\linewidth]{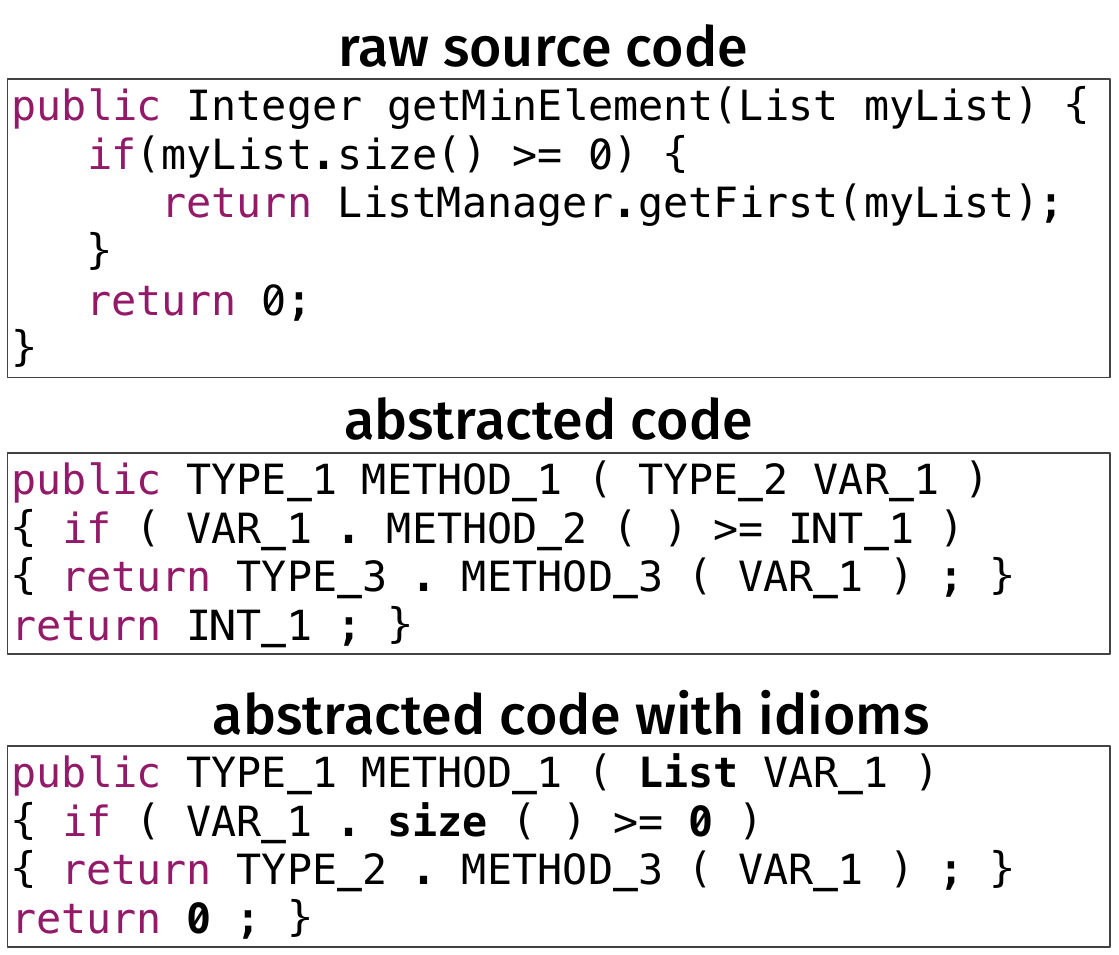}
	\caption{Abstraction process \cite{Tufano:tosem2019}}
	\label{fig:abstraction}
\end{figure}

To build this dataset, the authors mined $\sim$787k bug-fixing commits from GitHub, from which they extracted $\sim$2.3M BFPs. After that, the code of the BFPs is abstracted to make it more suitable for the NMT model (\ie to reduce the vocabulary of terms used in the source code identifiers and literals).  The abstraction process is depicted in \figref{fig:abstraction}. 

The top part of the figure represents the raw source code to abstract. The authors use a Java lexer and a parser to represent each method as a stream of tokens, in which Java keywords and punctuation symbols are preserved and the role of each identifier (\eg whether it represents a variable, method, \etc) as well as the type of a literal is discerned. 

IDs are assigned to identifiers and literals by considering their position in the method to abstract: The first variable name found will be assigned the ID of VAR\_1, likewise the second variable name will receive the ID of VAR\_2. This process continues for all identifiers as well as for the literals (\eg STRING\_X, INT\_X, FLOAT\_X). The output of this stage is the code reported in the middle of \figref{fig:abstraction} (\ie abstracted code). Since some identifiers and literals appear very often in the code (\eg variables \texttt{i}, \texttt{j}, literals \texttt{0}, \texttt{1}, method names such as \texttt{size}), those are treated as ``idioms'' and are not abstracted (see bottom part of \figref{fig:abstraction}, idioms are in bold). Tufano \etal consider as idioms the top $0.005\%$ frequent words in their dataset. During the abstraction a mapping between the raw and the abstracted tokens is maintained, thus allowing to reconstruct the concrete code from the abstract code generated by the model. 

The set of abstracted BFPs has been used to train and test the approach. The authors build two different sets, namely $BFP_{small}$, only including methods having a maximum length of 50 tokens (for a total of 58,350 instances), and $BFP_{medium}$, including methods up to 100 tokens (65,455). The model was able to correctly predict the patch for the buggy code in 9\% and 3\% of cases in the $BFP_{small}$ and $BFP_{medium}$ dataset, respectively. 

While other works have tackled the automatic bug-fixing problem, the approach by Tufano \etal has been tested on a variety of different bugs, rather than on specific types of bugs/warnings (\eg only single-line bugs are considered in \cite{Chen:2019}, while compilation failures are addressed in \cite{Mesbah:fse2019}). 

Thus, we picked it as representative DL technique for automatic bug-fixing and we use the two datasets by Tufano \etal \cite{Tufano:tosem2019} to fine-tune the T5 model for the ``automatic bug-fixing'' problem, comparing the achieved performance with the one reported in the original paper. 

\subsection{Injection of Code Mutants}

Brown \etal \cite{Brown:fse2017} were the first to propose a data-driven approach for generating code mutants, leveraging bug-fixes performed in software systems to extract syntactic-mutation patterns from the diffs of patches. Tufano \etal \cite{Tufano:icsme2019} built on this idea by presenting an approach using NMT to inject mutants representative of real bugs. The idea is similar to the previously described ``bug-fixing'' paper \cite{Tufano:tosem2019} with, however, the learning happening in the opposite direction. Indeed, given a bug-fixing commit, the input to the model is in this case the ``fixed method'' (\ie the method obtained after the bug-fixing activity) while the target is the buggy method (before the bug-fix). This allows the model to learn how to inject in a working code a mutant representative of real bugs. The applied methodology is the same described for the bug-fixing work \cite{Tufano:icsme2019}, including the abstraction process. 

This is, to date, the only DL-based technique for injecting code mutants. Thus, we use the dataset exploited by Tufano \etal \cite{Tufano:icsme2019} to fine-tune the T5 model for the problem of ``injecting code mutants'', comparing the achieved results with the ones reported in the original paper. Specifically, we reused their largest dataset, referred to as $GM_{ident}$ in the paper\footnote{A subset of this dataset named $GM_{ident-lit}$ has also been used in the original paper \cite{Tufano:icsme2019} to avoid including in the study bugs requiring the generation of previously unseen literals. We decided to test the T5 model on the most complex and complete dataset.}, featuring 92,476 training instances, 11,560 used for hyperparameter tuning (evaluation set), and 11,559 used for testing. On this data, the approach by Tufano \etal was able to correctly predict the bug to inject in 17\% of cases (1,991).

\subsection{Generation of Assert Statements in Test Methods}
Watson \etal \cite{Watson:icse2020} start from the work by Shamshiri \etal~\cite{Shamshiri:FSE'15}, who observed that tools for the automatic generation of test cases such as Evosuite \cite{evosuite}, Randoop \cite{randoop} and Agitar \cite{agitar} exhibit insufficiencies in the automatically generated assert statements. 

Thus, they propose ATLAS, an approach for generating syntactically and semantically correct unit test assert statements using NMT. To train ATLAS, the authors mined 2.5M test methods from GitHub with their corresponding {\tt assert} statement. For each of those test methods, they also identified the focal method, meaning the main production code method exercised by the test. A preprocessing of the dataset has been performed to remove all test methods longer than 1K tokens. Also, test methods requiring the synthesis of one or more unknown tokens for generating the appropriate assert statements have been removed. Indeed, if the required tokens cannot be found in the vocabulary of the test method they cannot be synthesized when the model attempts to generate the prediction. Finally, all duplicates have been removed from the dataset, leading to a final set of 158,096 Test-Assert Pairs (TAPs). Each method left in the dataset has then been abstracted using the same approach previously described by Tufano \etal \cite{Tufano:tosem2019}. However, in this case the authors experiment with two datasets, one containing raw source code and one abstracted code. ATLAS was able to generate asserts identical to the ones written by developers in 31.42\% of cases (4,968 perfectly predicted assert statements) when only considering the top-1 prediction, and 49.69\% (7,857) when looking at the top-5 in the abstracted dataset, while performance is lower on the raw dataset (17.66\% for top-1 and 23.33\% for top-5).

This is the only DL-based technique proposed in the literature to generate assert statements. We use the datasets by Watson \etal \cite{Watson:icse2020} to fine-tune our T5 model for the ``generation of assert statements'' problem, and compare the achieved performance with the one in the original paper. 

\subsection{Code Summarization}
Code summarization is one of the mainstream methods for automatic documentation of source code. The proposed summarization techniques fall into two categories: extractive \cite{Haiduc:wcre2010,Sridhara:icpc2011,Moreno:icpc2013,Rodeghero:icse17} and abstractive \cite{Sridhara:icse2011,McBurney:tse2016,Jiang:ASE'17,Hu:icpc2018,haque:2020}. The former create a summary of a code component which includes information extracted from the component being summarized, while the latter may include in the generated summaries information that is not present in the code component to document. DL techniques have been used to support the generation of abstractive summaries.

Hu \etal \cite{Hu:icpc2018} use a Deep Neural Network (DNN) to automatically generate comments for a given Java method. The authors mine $\sim$9k Java projects hosted on GitHub to collect pairs of $\langle$method, comment$\rangle$, where ``comment'' is the first sentence of the Javadoc linked to the method. These pairs, properly processed, are used to train and test the DNN. The authors assess the effectiveness of their technique by using the BLEU-4 score \cite{Papineni:2002}, showing the superiority of their approach with respect to the competitive technique presented in \cite{iyer:acl}.

Allamanis \etal \cite{Allamanis:2016} use attention mechanisms in neural networks to suggest a descriptive method name starting from an arbitrary snippet of code. Their approach can name a code snippet exactly as a developer would do in $\sim$25\% of cases. 

LeClair \etal \cite{LeClair:icse2019} present a neural model combining the AST source code structure and words from code to generate coherent summaries of Java methods. The approach, tested on 2.1M methods, showed its superiority as compared to the previous works by Hu \etal \cite{Hu:icpc2018} and Iyer \etal \cite{iyer:acl}.

The approach by Haque \etal \cite{haque:2020} is the most recent in the area of DL-aided source code summarization, and it is an improvement of the work by LeClair \etal \cite{LeClair:icse2019}. 

It still aims at documenting Java methods through an encoder-decoder architecture but, in this case, three inputs are provided to the model to generate the summary: (i) the source code of the method, as a flattened sequence of tokens representing the method; (ii) its AST representation; and (iii) the ``file context'', meaning the code of every other method in the same file. The authors show that adding the contextual information as one of the inputs substantially improves the BLEU score obtained by deep learning techniques. The dataset used in the evaluation is composed of 2.1M Java methods paired with summaries. We reuse this dataset for the fine-tuning of the T5 model for the code summarization problem, and compare its performance to the state-of-the-art approach proposed by Haque \etal \cite{haque:2020}.


\newcommand{\ReportDataset}[2]{$\mathit{#1}_{\mathit{#2}}$\xspace}

\newcommand{\BF}[1]{\ReportDataset{BF}{#1}}
\newcommand{\BFsmall}{\BF{small}}
\newcommand{\BFmedium}{\BF{medium}}

\newcommand{\AG}[1]{\ReportDataset{AG}{#1}}
\newcommand{\AGabs}{\AG{abs}}
\newcommand{\AGraw}{\AG{raw}}

\newcommand{\MG}[1]{\ReportDataset{MG}{#1}}
\newcommand{\MGident}{\MG{ident}}

\newcommand{\CS}{\ReportDataset{CS}{}}

\section{Multitask Learning for Code-related Tasks} \label{sec:approach}

\begin{figure*}
	\centering
	\includegraphics[width=\linewidth]{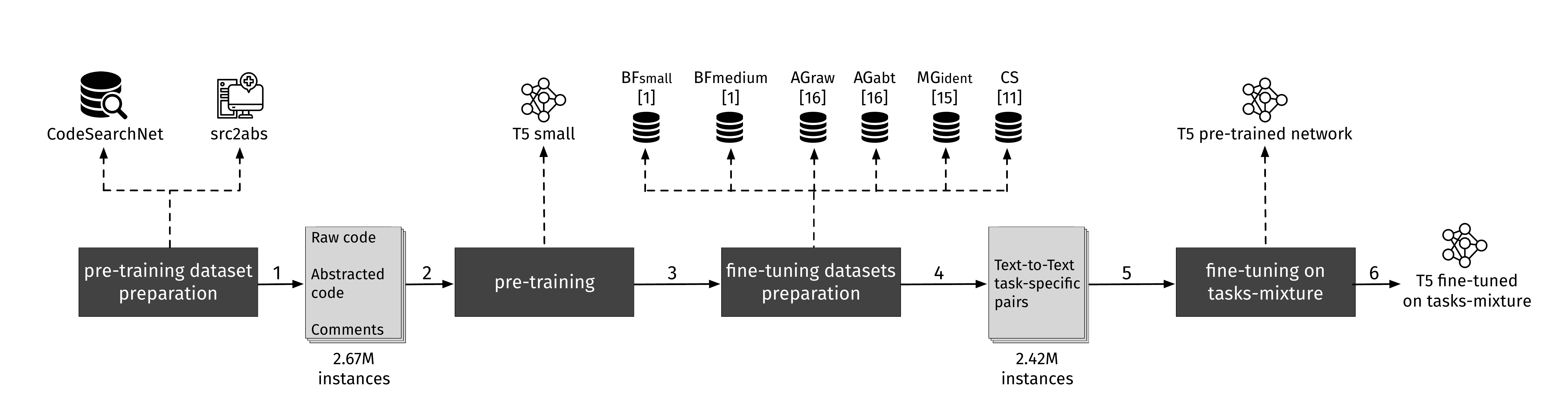}
	\caption{\revised{Overview of the approach used to pre-train and fine-tune the T5 model.}}
	\label{fig:approach}
\end{figure*}

The T5 model was introduced by Raffel \etal \cite{raffel2019exploring} to support multitask learning in the domain of NLP. This approach is based on two phases: \textit{pre-training}, which allows defining a shared knowledge-base useful for a large class of sequence-to-sequence tasks, and \textit{fine-tuning}, which specializes the model to specific tasks of interest.
In this section, we first provide basic information about the T5 model (refer to \cite{raffel2019exploring} for a detailed explanation of the architecture). Then, we explain how we adapted it to the software engineering domain, with the goal of supporting the four tasks previously described: \textit{automatic bug-fixing}, \textit{generation of assert statements in test methods}, \textit{code summarization}, and \textit{injection of code mutants}. Such a process is depicted in \figref{fig:approach}. Finally, we describe the hyperparameter tuning of the model and the adopted decoding strategy.

\subsection{T5 in a Nutshell}
The T5 model is based on the transformer model architecture \cite{vaswani2017attention} that allows to handle a variable-sized input using stacks of self-attention layers \cite{cheng2016long} instead of RNNs or CNNs.  
When an input sequence is provided, it is mapped to a sequence of embeddings that is passed into the encoder. 

The \textit{encoders} are all identical in structure and each one is comprised of two subcomponents: a \textit{self-attention layer} followed by a small \textit{feed-forward network}. Layer normalization \cite{ba2016layer} is applied to the input of each subcomponent while a residual skip connection \cite{he2016deep} adds each input of the subcomponent to its output. Dropout \cite{srivastava2014dropout} is applied within the feed-forward network, on the skip connection, on the attention weights, and at the input and output of the entire stack. 
The \textit{decoders} work similarly to the encoders: Each self-attention layer is followed by an additional attention mechanism that attends to the output of the encoder.

The output of the final decoder block is fed into a dense layer with a softmax output, to produce the output probabilities over the vocabulary. Differently from the generic transformer model, the T5 model \cite{raffel2019exploring} uses a simplified form of position embeddings, where each embedding is a scalar that is added to the corresponding logit used for computing the attention weights. As pointed out by the authors, for efficiency they also share the position embedding parameters across all layers.

The T5, in particular, and a transformer model, in general, offer two main advantages over other state-of-the-art models: (i) it is more efficient than RNNs since it allows to compute the output layers in parallel, and (ii) it is able to detect hidden and long-ranged dependencies among tokens, without assuming that nearest tokens are more related than distant ones. This last property is particularly relevant in code-related tasks since a variable declaration may be distant from its usage.

Five different versions of T5 have been proposed \cite{raffel2019exploring}: \textit{\textit{small}}, \textit{\textit{base}}, \textit{\textit{large}}, \textit{\textit{3 Billion}}, and \textit{\textit{11 Billion}}. These variants differ in terms of complexity, with the smaller model (T5\textsubscript{\textit{small}}) having 60M parameters against the 11B of the largest one (T5\textsubscript{\textit{11B}}). As acknowledged by the authors \cite{raffel2019exploring}, even if the accuracy of the most complex variants are higher than the less complex models, the training complexity increases with the number of parameters. Considering the available computational resources, in our work we decided to use the simplest T5\textsubscript{\textit{small}} model. We expect the results achieved in our study to be a lower bound for the performance of a T5-based model. Nevertheless---as reported in Section \ref{sec:results}---the T5\textsubscript{\textit{small}} model is still able to outperform state-of-the-art approaches.

The T5\textsubscript{\textit{small}} architecture is characterized by six blocks for encoders and decoders. The feed-forward networks in each block consist of a dense layer with an output dimensionality ($d_{ff}$) of 2,048. The \textit{key} and \textit{value} matrices of all attention mechanisms have an inner dimensionality ($d_{kv}$) of 64, and all attention mechanisms have eight heads. All the other sub-layers and embeddings have a dimensionality ($d_{model}$) of 512. 

\subsection{Pre-training of T5}\label{subsec:training_strategy}

In the \textit{pre-training} phase we use a self-supervised task similar to the one used by Raffel \etal \cite{raffel2019exploring}, consisting of masking tokens in natural language sentences and asking the model to guess the masked tokens. Differently, we did not perform the pre-training by only using natural language sentences, since all the tasks we target involve source code. Thus, we use a dataset composed of both (technical) natural language (\ie code comments) and source code. To obtain the dataset for the pre-training we start from the CodeSearchNet dataset \cite{husain2019codesearchnet}, which provides 6M functions from open-source code. We only focus on the $\sim$1.5M methods written in Java, since the four tasks we aim at supporting are all related to Java code. Then, since for three of the four tasks we support (\ie \emph{automatic bug-fixing} \cite{Tufano:tosem2019}, \emph{generation of assert statements} \cite{Watson:icse2020}, and \emph{injection of code mutants} \cite{Tufano:icsme2019}) the authors of the original papers used an abstracted version of source code (see \secref{sec:related}), we used the {\tt src2abs} tool by Tufano \cite{Tufano:tosem2019} to create an abstracted version of each mined Java method. 

Note that, since the tool was run on Java methods in isolation (\ie without providing it the whole code of the projects they belong to), {\tt src2abs} raised a parsing error in $\sim$600k of the $\sim$1.5M methods (due \eg to missing references), leaving us with $\sim$900k abstracted methods. We still consider such a dataset as sufficient for the pre-training. 

The CodeSearchNet dataset does also provide, for a subset of the considered Java source code methods, the first sentence in their Javadoc. We extracted such a documentation using the {\tt docstring\_tokens} field in CodeSearchNet, obtaining it for 499,618 of the considered methods. We added these sentences to the pre-training dataset. This whole process resulted in a total of 2,984,627 pre-training instances, including raw source code methods, abstracted methods, and code comment sentences. Finally, in the obtained dataset there could be duplicates between (i) different raw methods that become equal once abstracted, and (ii) comments re-used across different methods. Thus, we remove duplicates, obtaining the final set of  2,672,450 instances reported in \tabref{tab:pretrain_dataset}. This is the dataset we use for pre-training the T5 model, using the BERT-style objective function Raffel \etal used in their final experiments and consisting of randomly masking 15\% of tokens (\ie words in comments and code tokens in the raw and abstracted code). 

\begin{table}[h]
	\centering
	\begin{tabular}{lr}
        \toprule
		\textbf{Data sources}                        & \textbf{Instances}  \\
        \midrule
		Source code                  & 1,569,773           \\
		Abstracted source code        &   766,129           \\
		Technical natural language    &   336,548           \\
        \midrule                                     
		\textbf{Total} & 2,672,450           \\
		\bottomrule
	\end{tabular}
    \caption{Datasets used for the pre-training of T5.}
	\label{tab:pretrain_dataset}
\end{table}

Finally, since we pre-train the model on a software-specific dataset, we needed to create a new vocabulary to accommodate the tokens in our dataset. For this reason, we created a new \emph{SentencePiece} model \cite{DBLP:journals/corr/abs-1808-06226} (\ie a tokenizer for neural text processing) by using the entire pre-training dataset.

\begin{table}
	\centering
	\resizebox{.5\textwidth}{!}{
		\begin{tabular}{llrrr}
			\toprule
			\textbf{Task} & \textbf{Dataset}                  & \textbf{Evaluation-set} & \textbf{Training-set} & \textbf{Test-set}  \\
			\midrule                                          
			\textbf{Bug Fixing}
                          & \BFsmall \cite{Tufano:tosem2019}  &   5,835                 & 46,680                &  5,835  \\      
                          & \BFmedium \cite{Tufano:tosem2019} &   6,546                 & 52,364                &  6,545  \\

			\textbf{Mutant Generation}
                          & \MGident \cite{Tufano:icsme2019}  &  11,560                 & 92,476                &  11,559 \\
                          
			\textbf{Assert Generation}
                          & \AGabs \cite{Watson:icse2020}     &  15,809                 & 126,477               &  15,810 \\
                          & \AGraw \cite{Watson:icse2020}     &  18,816                 & 150,523               &  18,815 \\
                          
            \textbf{Code Summarization}
                          & \CS \cite{haque:2020}             & 104,272                 & 1,953,940             &  90,908 \\
                          
			\midrule                                          
			\textbf{Total}                    &               & 162,838                 & 2,422,460             & 149,472 \\
			\bottomrule
		\end{tabular}
	}
	\caption{Task-specific datasets used for fine-tuning T5.}
		\label{tab:finetuning_datasets}
	\vspace{-0.2cm}	
	\end{table}

\subsection{Fine-tuning of T5}\label{subsec:finetuning}
We use a slightly modified version of the multi-task learning approach used by Raffel \etal \cite{raffel2019exploring}: we fine-tune the model on a mixture of tasks instead of performing fine-tuning for each single task. 

We do this because of the relatively small size of the specialized datasets available. \tabref{tab:finetuning_datasets} reports summary characteristics of the datasets we use for each task.
Also, for each task we have to provide a consistent framing of the input that allows the model to recognize the tasks that should be performed given an input sequence of tokens. We use a special token sequence indicating the task at hand (\eg ``generate small patch'' for \BFsmall, followed by the token ``:'' and by the input required by the task.

\subsubsection{Datasets Used for Fine-Tuning}

In the following, we describe the details of the datasets we use for \textit{fine-tuning} the model for the four targeted tasks.

\textbf{Automatic Bug Fixing (BF)}. 
We use the dataset by Tufano \etal \cite{Tufano:tosem2019} composed by triplets $\mathit{BF}_m = \langle  m_\mathit{b}, m_\mathit{f}, M \rangle$, where $m_\mathit{b}$ and $m_\mathit{f}$ are the abstracted version of the buggy and fixed version of Java method, respectively, and $M$ represents the mapping between the abstracted tokens and the raw code tokens (\eg \texttt{VAR\_1} $\rightarrow$ \texttt{webServerPort}), which allows to track back the output of the model to source code.
The triplets refer to methods with at most 100 tokens and they are split into two sub-datasets: (i) the \textit{small} version, containing methods with up to 50 tokens, and a \textit{medium} version, with methods with at most 100 tokens.
We train the model to predict the fixed versions, $m_\mathit{f}$, given the buggy versions, $m_\mathit{b}$. Given the presence of two datasets, we divide the BF task in two sub-tasks, \BFsmall and \BFmedium, depending on the size of the method \cite{Tufano:tosem2019}.

\textbf{Injection of Code Mutants (MG)}.
For the MG task we exploited one of the two datasets provided by Tufano \etal \cite{Tufano:icse2019}: \MGident and \MG{ident-lit}. In both datasets each instance is represented by a triple $\langle  m_\mathit{f}, m_\mathit{b}, M \rangle$, where, similarly to the BF datasets, $m_\mathit{b}$ and $m_\mathit{f}$ are the buggy and fixed version of the snippet, respectively, and $M$ represents the mapping between the abstracted tokens and the code tokens. 

The first dataset (\MGident) represents the most general (and challenging) case, in which the mutated version, $m_\mathit{b}$, can also contain new tokens (\ie identifiers, types, or method names) not contained in the version provided as input ($m_\mathit{f}$). \MG{ident-lit}, instead, only contains samples in which the mutated version contains a subset of the tokens in the non-mutated code. In other words, \MG{ident-lit} represents a simplified version of the task. For this reason, we decided to focus on the most general scenario and we only use the \MGident dataset. 

\textbf{Generation of Assertions in Test Methods (AG)}.
For the AG task we used the dataset provided by Watson \etal \cite{Watson:icse2020} containing triplets $\langle T, TM_n, A\rangle$, where $T$ is a given test case, $TM_n$ is the \textit{focal} method tested by $T$, \ie the last method called in $T$ before the assert \cite{trace_link}, and $A$ is the assertion that must be generated (output). For such a task, we use two versions of the dataset: \AGraw, which contains the raw source code for the input ($T + TM_n$) and the output ($A$), and \AGabs, which contains the abstracted version of input and output, \ie $src2abs(T + TM_n)$ and $src2abs(A)$, respectively. These are the same datasets used in the original paper.

\textbf{Code Summarization (CS)}.
For code summarization, we exploited the dataset provided by Haque \etal \cite{haque:2020} containing 2,149,120 instances, in which each instance is represented by a tuple $\langle S, A_S, C_S, D \rangle $, where $S$ represents the raw source code of the method, $A_S$ is its AST representation, $C_S$ is the code of other methods in the same file, and $D$ is the summary of the method, \ie the textual description that the model should generate \cite{haque:2020}. For this specific task, we consider a variation of the original dataset to make it more coherent with the performed pre-training. In particular, since in the pre-training we did not use any AST representation of code, we decided to experiment with the T5 model in a more challenging scenario in which only the raw source code to summarize (\ie $S$) is available to the model. Therefore, the instances of our dataset are represented by tuples $\langle S, D \rangle$: We train our model to predict $D$ given only $S$.


\subsubsection{Data Balancing}
The datasets we use for fine-tuning have different sizes, with the one for code summarization dominating the others. This could result in an unbalanced effectiveness of the model on the different tasks. In our case, the model could become very effective in summarizing code and less in the other three tasks. However, as pointed out by Arivazhagan \etal \cite{DBLP:journals/corr/abs-1907-05019}, there is no free lunch in choosing the balancing strategy when training a multi-task model, with each strategy having its pros and cons (\eg oversampling of less represented datasets negatively impacts the performance of the most representative task). For this reason, while fine-tuning, we decided not to perform any particular adaptation of our training set, following the true data distribution when creating each batch: We sample instances from the tasks in such a way that each batch during the training has a proportional number of samples accordingly to the size of the training dataset.

\subsection{Decoding Strategy}
Given the values of the output layer, different decoding strategies can be used to generate the output token streams. 

T5 allows to use both \textit{greedy decoding} and \textit{Beam-search}. When generating an output sequence, the greedy decoding selects, at each time step $t$, the symbol having the highest probability. The main limitation of greedy decoding is that it only allows the model to generate one possible output sequence (\eg one possible bug fix) for a given input sequence (\eg the buggy method).

Beam-search is an alternative decoding strategy previously used in many DL applications \cite{DBLP:journals/corr/abs-1211-3711, boulanger2013audio, DBLP:journals/corr/BahdanauCB14, Raychev:2014:CCS:2594291.2594321}. Unlike greedy decoding, which keeps only a single hypothesis during decoding, beam-search of order $K$, with $K > 1$, allows the decoder to keep $K$ hypotheses in parallel: At each time step $t$, beam-search picks the $K$ hypotheses (\ie sequences of tokens up to $t$) with the highest probability, allowing the model to output $K$ possible output sequences.

We used Beam-search to provide several output sequences given a single input, and report results with different $K$ values. It is worth noting that having a large $K$ increases the probability that one of the output sequences is correct, but, on the other hand, it also increases the cost of manually analyzing the output for a user (\ie a developer, in our context).

\subsection{Hyperparameter Tuning}
For the \textit{pre-training} phase, we use the default parameters defined for the T5 model \cite{raffel2019exploring}. Such a phase, indeed, is task-agnostic, and hyperparameter tuning would provide limited benefits. Instead, we tried different learning rate strategies for the \textit{fine-tuning} phase. Especially, we tested four different learning rates: (i) \textit{Constant Learning Rate} (C-LR): the learning rate is fixed during the whole training (we use $LR = 0.001$, \ie the value used in the original paper \cite{raffel2019exploring}); (ii) \textit{Inverse Square Root Learning Rate} (ISR-LR): the learning rate decays as the inverse square root of the training step (the same used for pre-training by Raffel \etal); (iii) \textit{Slanted Triangular Learning Rate \cite{howard2018universal}} (ST-LR): the learning rate first linearly increases and then linearly decays to the starting learning rate; 
(iv) \textit{Polynomial Decay Learning Rate} (PD-LR): the learning rate decays polynomially from an initial value to an ending value in the given decay steps. 

\tabref{tab:hyperparameter:types} reports the specific parameters we use for each scheduling strategy: the values are the default ones reported in the papers that introduced them.

\begin{table}[h]
\vspace{0.4cm}
 \centering
 \begin{tabular}{ll}
  \toprule
  \textbf{Learning Rate Type}   & \textbf{Parameters}\\
  \midrule
  Constant            & $\mathit{LR} = 0.001$        \\
  Inverse Square Root 
                      & $\mathit{LR}_{\mathit{starting}} = 0.01$ \\
                      & $\mathit{Warmup} = 10,000$ \\
  Slanted Triangular  
                      & $\mathit{LR}_{\mathit{starting}} = 0.001$ \\
                      & $\mathit{LR_{\mathit{max}}} = 0.01$\\
                      & $\mathit{Ratio} = 32$\\
                      & $\mathit{Cut} = 0.1$\\
  Polynomial Decay    
                      & $\mathit{LR}_{\mathit{starting}} = 0.01$\\
                      & $\mathit{LR}_{\mathit{end}} = 1\mathrm{e}{-06}$\\
                      & $\mathit{Power} = 0.5$\\
  \bottomrule
 \end{tabular}
 \caption{Learning-rates tested for hyperparameter tuning.}
 \label{tab:hyperparameter:types}
\end{table}

We pre-train the model for a total of 100k steps in the four configurations on the whole pre-training set and we test it on the evaluation sets of the datasets provided by the original papers we compare with. 

We compute the following metrics: for BF and AG, we compute the percentage of perfect predictions achieved with the greedy decoding strategy (Accuracy@1); for MG, we compute the BLEU score \cite{Papineni:2002}; for CS, we compute BLEU-A, the geometric average of the BLEU-\{1,2,3,4\} scores \cite{Papineni:2002}. Basically, for each task we adopt one of the evaluation metrics used in the original paper (details about these metrics are provided in \secref{sub:metrics}). We report in \tabref{tab:hyperparameter:results} the achieved results (in bold the learning rate obtaining the best performance for each metric/dataset). As it can be noticed, the \emph{Slanted Triangular Learning Rate (ST-LR)} allows to achieve the best performance in most of the cases. For this reason, we decided to use this particular learning rate in our model.

Several other hyperparameters could have been tuned. Given the high computational cost to train the model ($\sim$343 hours on a colab \cite{colab} instance with 8 tpu cores and 35.5GB of RAM), we did not manage to perform a comprehensive hyperparameter tuning. The high dimensionality of the model, indeed, makes hyperparameter tuning not very cost-effective: we preferred to use the computational power available to increase the number of steps for training the model.

\begin{table}
    \centering
    \resizebox{\columnwidth}{!}{%
    \begin{tabular}{llrrrr}
    \toprule
        \textbf{Dataset}  & \textbf{Metric}    & \textbf{C-LR}   & \textbf{ST-LR}   & \textbf{ISQ-LR} & \textbf{PD-LR} \\
        \midrule
        \BFsmall \cite{Tufano:tosem2019}   
                          & Accuracy@1         &  6.9\%          & \textbf{13.2\%}  & 11.0\%          & 0.27\%         \\

        \BFmedium \cite{Tufano:tosem2019}                                                                             
                          & Accuracy@1         &  2.9\%          & \textbf{5.5\%}   &  3.3\%          &  0.0\%         \\
        \MGident \cite{Tufano:icsme2019}                                                                                     
                          & BLEU-A             & 75.6\%          & \textbf{78.2\%}  & 77.7\%          & 12.0\%         \\  

        \AGabs \cite{Watson:icse2020}                                                                                        
                          & Accuracy@1         & 33.7\%          & 39.7\%           & \textbf{39.8\%} & 2.0\%          \\

        \AGraw \cite{Watson:icse2020}                                                                                        
                          & Accuracy@1         & 48.9\%          & \textbf{57.6\%}  & 56.7\%          & 2.1\%          \\

        \CS \cite{haque:2020}                                                                                                
                          & BLEU-A             & 23.3\%          & 23.6\%           & \textbf{24.3\%} &  3.4\%         \\
        \midrule
        \multicolumn{2}{l}{\textbf{\# Best Results}}
                                               & 0               & \textbf{4}       & 2               & 0 \\
        \bottomrule
    \end{tabular}
   }
    \caption{Hyperparameter tuning results.}
    \label{tab:hyperparameter:results}
\end{table}


\newcommand{\RQ}[1]{RQ\textsubscript{#1}}
\newcommand{\Shared}[1]{$\mathit{Shared}_{\mathit{#1}}$}
\newcommand{\OnlyOurs}[1]{$\mathit{OnlyT5}_{\mathit{#1}}$}
\newcommand{\OnlyBaseline}[1]{$\mathit{OnlyBL}_{\mathit{#1}}$}
\section{Study Design} \label{sec:design}

The \textit{goal} of our study is to understand whether \emph{multi-task learning}, in general, and a T5-based model, in particular, is suitable for automating code-related tasks. The \textit{context} is represented by the datasets introduced in \secref{sec:related}, \ie the ones by Tufano \etal for  bug fixing \cite{Tufano:tosem2019} and injection of mutants \cite{Tufano:icsme2019}, by Watson \etal for assert statement generation \cite{Watson:icse2020}, and by Haque \etal for code summarization \cite{haque:2020}. 

Our study is steered by the following research question: \emph{Is \revised{the T5 model} suitable for code-related tasks such as automatic bug fixing, injection of mutants, assert statement generation and code summarization?}
%
%

%

\subsection{Experimental Procedure}
\label{sub:metrics}
We use the model we trained and tuned as we specified in \secref{sec:approach} and we run it on the test sets provided in the previously described datasets. Our baselines are the state-of-the-art models described in \secref{sec:related}. For each task and dataset, we compare the results achieved by our model with the results reported in the original papers. 

We use different metrics for the different tasks, depending on the metrics reported in the papers that introduced our baselines.
\tabref{tab:design} reports the baselines and metrics used to evaluate the results for each task, that we define below.

\begin{table}
 \centering
 \begin{tabular}{lllll}
  \toprule
  \textbf{Task}     & \textbf{Baseline}        & \textbf{Accuracy@K}      & \textbf{BLEU-n}         & \textbf{ROUGE LCS} \\
  \midrule
  BF                & \cite{Tufano:tosem2019}  & $\{1, 5, 10, 25, 50\}$   &   -                      &         -       \\
  MG                & \cite{Tufano:icsme2019}  & $\{1\}$                  & $\{A\}$        &       -         \\
  AG                & \cite{Watson:icse2020}   & $\{1, 5, 10, 25, 50\}$   &   -                       &   -             \\
  CS                & \cite{haque:2020}        &  -                        & $\{1, 2, 3, 4, A\}$                 & $\{P, R, F\}$  \\
  \bottomrule
 \end{tabular}
 \caption{Baselines and evaluation metrics for the tasks.}
 \label{tab:design}
\end{table}

\textbf{Accuracy@K} measures the percentage of cases (\ie instances in the test set) in which the  sequence predicted by the model equals the oracle sequence (\ie perfect prediction). Since we use beam-search, we report the results for different $K$ values (\ie 1, 5, 10, 25, and 50), as done in \cite{Tufano:tosem2019} (BF) and \cite{Watson:icse2020} (AG). Tufano \etal \cite{Tufano:icse2019} do not report results for $K > 1$ for the MG task. Thus, we only compare the results with $K = 1$.

\textbf{BLEU score} (Bilingual Evaluation Understudy) \cite{Papineni:2002} measures how similar the candidate (predicted) and reference (oracle) texts are. Given a size $n$, the candidate and reference texts are broken into \textit{n}-grams and the algorithm determines how many \textit{n}-grams of the candidate text appear in the reference text. The BLEU score ranges between 0 (the sequences are completely different) and 1 (the sequences are identical).
We use different BLEU-\textit{n} scores, depending on the ones used in the reference paper of the baseline. For the CS task, we report BLEU-\{1, 2, 3, 4\} and their geometric mean (\ie BLEU-A); for the MG task we only report BLEU-A.

\textbf{ROUGE} (Recall-Oriented Understudy for Gisting Evaluation) is a set of metrics for evaluating both automatic summarization of texts and machine translation techniques in general \cite{lin2004rouge}. ROUGE metrics compare an automatically generated summary or translation with a set of reference summaries (typically, human-produced). We use the ROUGE LCS metrics based on the Longest Common Subsequence for the CS task \cite{haque:2020}. Given two token sequences, $X$ and $Y$, and their respective length, $m$ and $n$, it is possible to compute three ROUGE LCS metrics: $R$ (recall), computed as $\frac{LCS(X, Y)}{m}$, $P$ (precision), computed as $\frac{LCS(X, Y)}{n}$, and F (F-measure), computed as the harmonic mean of $P$ and $R$.

Besides such effectiveness metrics, we also perform an additional analysis: 
we compute the \textit{inference time}, \ie the time needed to run the model on a given input. We run such an experiment on a laptop equipped with a 2.3GHz 8-core 9th-generation Intel Core i9 and 16 GB of RAM. We do this for different beam search sizes, with $K \in \{1, 5, 10, 25, 50\}$. For each $K$, we report the average inference time on all the instances of each task. This allows understanding the \textit{efficiency} of the model and to what extent it can be used in practice.

Finally, for each task, we also compute the complementarity between T5 and the baseline approach. For each dataset $d$ we consider and the related baseline approach $\mathit{BL}_d$, we first define the sets of perfect predictions obtained by the two approaches $\mathit{PP}_{\mathit{T5}_d}$ and $\mathit{PP}_{\mathit{BL}_d}$ with a fixed beam size $K=1$. 

Then, we compute three metrics:

\footnotesize
$$
  \mathit{Shared}_{\mathit{d}} = \frac{|\mathit{PP}_{\mathit{T5}_d} \cap \mathit{PP}_{\mathit{BL}_d}|}{|\mathit{PP}_{\mathit{T5}_d} \cup \mathit{PP}_{\mathit{BL}_d}|}
$$

\noindent\begin{minipage}{.5\linewidth}
$$
  \mathit{OnlyT5}_{\mathit{d}} = \frac{|\mathit{PP}_{\mathit{T5}_d} \setminus \mathit{PP}_{\mathit{BL}_d}|}{|\mathit{PP}_{\mathit{T5}_d} \cup \mathit{PP}_{\mathit{BL}_d}|}
$$
\end{minipage}%
\begin{minipage}{.5\linewidth}
$$
  \mathit{OnlyBL}_{\mathit{d}} = \frac{|\mathit{PP}_{\mathit{BL}_d} \setminus \mathit{PP}_{\mathit{T5}_d}|}{|\mathit{PP}_{\mathit{T5}_d} \cup \mathit{PP}_{\mathit{BL}_d}|}
$$
\smallskip
\end{minipage}
\normalsize

\Shared{d} measures the percentage of perfect predictions shared between the two compared approaches, while \OnlyOurs{d} and \OnlyBaseline{d} measure the percentage of cases in which the perfect prediction is only achieved by T5 or the baseline, respectively, on the dataset $d$.



\begin{figure*}
	\centering
	\includegraphics[width=\linewidth]{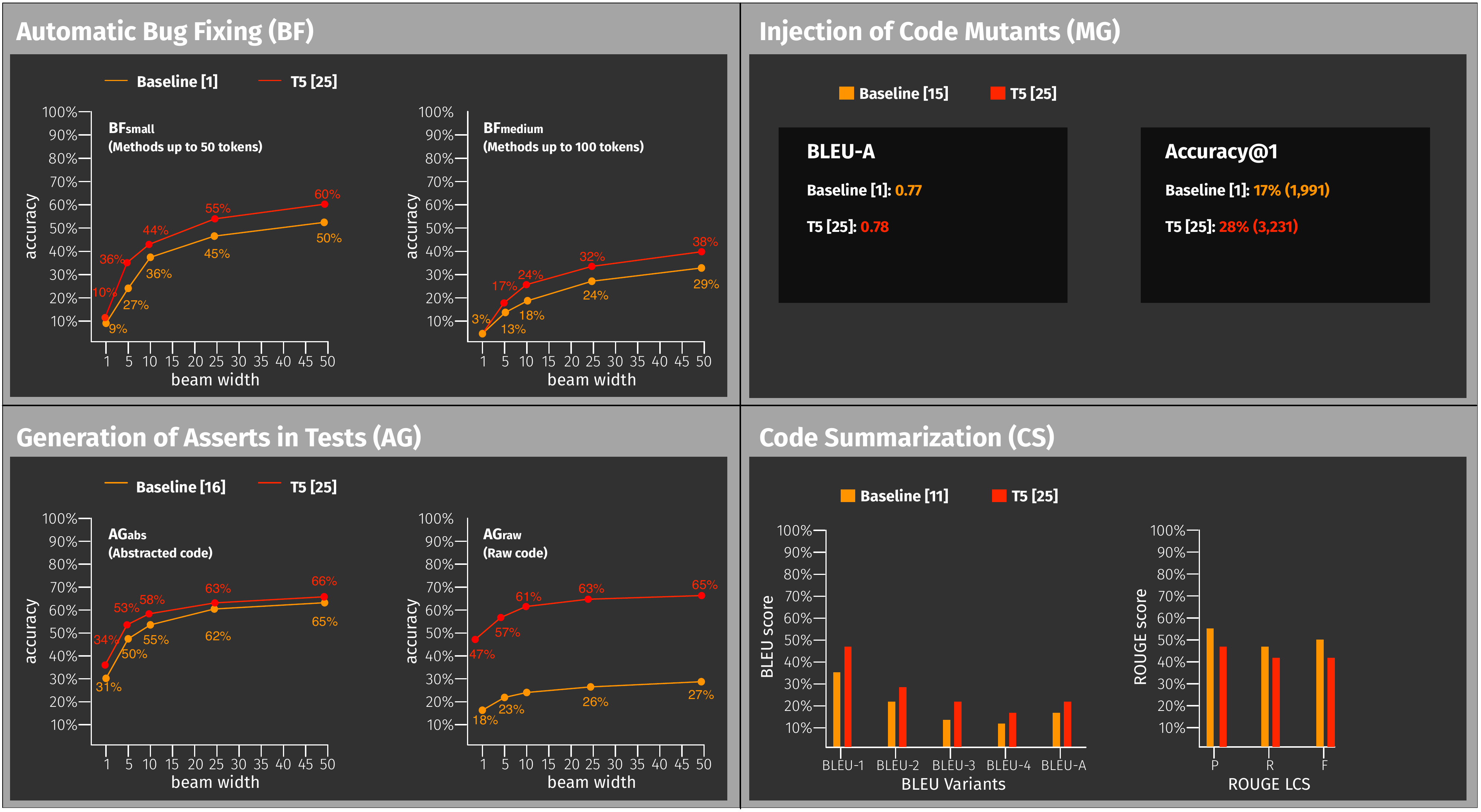}
	\caption{Performance of the T5 model against the experimented baselines.}
	\label{fig:results}
\end{figure*}

\newcommand{\about}{$\sim$}
\section{Results Discussion} \label{sec:results}

We report a summary of the results achieved by T5 (in red) and by the respective baselines (in orange) for the four tasks we consider (\ie BF, MG, AG, and CS) in \figref{fig:results}. We also show the inference times for all the tasks and the overlap metrics between T5 and the experimented baselines in \tabref{tab:inference_time} and \tabref{tab:overlap}, respectively. We discuss the results task by task below.

\begin{table}[h]
	\caption{Inference time with different beam size values.}
	\label{tab:inference_time}
	\resizebox{\linewidth}{!}{
		\begin{tabular}{l l l l l l l}
			\toprule
			$K$ & \BFsmall & \BFmedium & \MGident & \AGabs & \AGraw & \CS  \\
			\midrule                        
			1   & 0.41     & 1.84      & 0.31     & 0.35   & 0.36   & 0.12 \\
			5   & 0.62     & 1.13      & 0.54     & 0.79   & 0.66   & 0.17 \\
			10  & 0.72     & 1.55      & 0.62     & 1.17   & 1.20   & 0.24 \\
			25  & 1.30     & 3.35      & 1.13     & 2.45   & 2.66   & 0.40 \\
			50  & 2.16     & 5.31      & 2.04     & 4.82   & 4.96   & 0.74 \\
			\bottomrule
		\end{tabular}
	}
\end{table}

\begin{table}
	\centering
	\caption{\revised{Overlap metrics for correct predictions generated by the T5 model and the baselines.}}
	\label{tab:overlap}
	\begin{tabular}{l r r r}
		\toprule
		\emph{Dataset} ($d$) & \Shared{d} & \OnlyOurs{d} & \OnlyBaseline{d} \\
		\midrule
		\BFsmall             & 37.67\%    & 36.52\%      & 25.81\%          \\
		\BFmedium            & 28.78\%    & 36.06\%      & 35.16\%          \\
		\MGident             & 41.03\%    & 46.65\%      & 12.32\%          \\
		\AGabs               & 39.78\%    & 36.19\%      & 24.03\%          \\
		\AGraw               & 11.68\%    & 84.30\%      & 4.02\%           \\
		\CS                  & 4.97\%     & 93.46\%      & 1.57\%           \\
		\bottomrule
	\end{tabular}
\end{table}

%

\subsection{Automatic Bug Fixing (BF)}
When using T5 for automatically fixing bugs, the accuracy achieved using a greedy decoding strategy ($K = 1$) is very similar to the one achieved by the baseline on both the datasets we consider, \ie \BFsmall and \BFmedium. While on the first one there is a 1\% improvement, on the other the results are exactly the same. However, when increasing the beam size, the difference becomes larger: on \BFsmall the improvement ranges between 8-10\%, while on \BFmedium it is lower, and it ranges between 4-9\%. In general, it can be noticed that the improvement margin is constant.

The time needed to generate a fix depends on the dataset, \ie on the number of tokens of the input. If we use the \BFsmall dataset, the average inference time ranges between 0.41s ($K=1$) and 2.16s ($K=50$), while it is larger on the \BFmedium dataset (1.84s for $K=1$ and 5.31s for $K=50$).

There is a considerable overlap between the perfect predictions done by the two approaches (see \tabref{tab:overlap}): \about 38\% of perfect predictions on \BFsmall and \about 29\% on \BFmedium are shared by the two techniques. 

The remainder are perfect predictions only with T5 (\about 36\% on \BFsmall and \about 36\% on \BFmedium) or only with the baseline (\about 26\% on \BFsmall and \about 35\% on \BFmedium). This indicates that the two approaches are complementary for the BF task suggesting that, even if T5 was not able to fix some bugs, it is still possible to automatically fix (a subset of) such bugs with a specialized ML-based approach. This recalls the need to further enrich the architecture of a transfer learning method with the goal of further improving its ability to exploit the knowledge acquired on specific tasks.



\subsection{Injection of Code Mutants (MG)}
Looking at \figref{fig:results}, we can observe that using T5 to generate mutants allows to obtain much more accurate results than the baseline, with the Accuracy@1 improving by 11\%, with 1,240 additional perfect predictions (+62\% as compared to the baseline). The average BLEU score improves by \about 0.01 on top of the very good results already obtained by the baseline (\ie 0.77). Minor improvements in BLEU score can still indicate major advances in the quality of the generated solutions\cite{googleblogentry}.

As for the inference time (\tabref{tab:inference_time}), we observed similar results compared to the BF task on the \BFsmall dataset: with $K = 1$, the average inference time is 0.31s, while for $K = 50$ it is 2.04s. We do not report perfect predictions at $K = 50$ since those were not reported in the original paper \cite{Tufano:icsme2019}.

Similarly to BF, also for MG the percentage of shared perfect predictions (\tabref{tab:overlap}) is quite high (\about 41\%) with, however, T5 being the only one generating \about 46\% of perfect predictions as compared to the \about 12\% of the baseline approach. 

%

\subsection{Generation of Assertions in Test Methods (AG)}
T5 achieves very similar results compared to the baseline on the \AGabs (see \figref{fig:results}): when abstracting the tokens, both approaches achieve very similar levels of accuracy, and such values are reasonably high with the increase of $K$ (\eg they both achieve 65\% accuracy with $K = 50$). However, when using the more challenging non-abstracted dataset \AGraw, T5 allows to achieve much better results: it achieves a 29\% higher accuracy with $K = 1$, while for larger $K$ values the gap in performance ranges between 35-38\%. The most interesting result, however, is that T5 achieves similar results both with and without abstraction, with the the Accuracy@1 being higher when considering \AGraw then when considering \AGabs. The fact that T5 is capable of handling raw source code makes its usage more straightforward compared to the baseline: it does not need pre- and post-processing steps for such a task.

Assert generation is very fast for low values of $K$ (0.36s for both the datasets with $K = 1$), while it gets much slower for higher values of $K$, at a higher rate compared to other tasks (4.82s for \AGabs and 4.96s for \AGraw with $K = 50$).

In terms of overlap, we found a trend similar to BF on \AGabs: we have \about 40\% of perfect predictions shared between the two approaches, while the remainder instances are distributed between the ones only predicted by T5 (\about 36\%) and the ones only predicted by the baseline (\about 24\%). 

There is, instead, a small overlap on the \AGraw dataset: only \about 12\% of the instances are perfectly predicted by both the approaches, with \about 84\% of them correctly predicted only by T5. 

%
%

\subsection{Code Summarization (CS)}
On this task, T5 achieves a substantial increase in BLEU score as compared to the baseline. When considering the average BLEU (BLEU-A), the improvement is of $\sim$5\%. On the other hand, it can be noticed that the ROUGE-LCS scores achieved when using T5 are lower than the ones achieved by the baseline ($\sim$6\% lower on the F-measure score). Thus, looking at these metrics, there is no clear winner, but T5 seems to be at least comparable to the baseline. To have something easier to interpret, we compared the two approaches in terms of the number of perfect predictions they generate, despite the fact that such a metric was not used in the original paper \cite{haque:2020}. This means counting the comments generated by a technique that are exactly equal to the ones manually written by humans. T5 managed to generate 11.4\% of perfect predictions (10,401 instances) against the 3.4\% (3,048) of the baseline technique (over 3 $\times$ better). 

Code summarization is the fastest task to complete for T5: it takes only 0.12s for $K = 1$ and 0.74s for $K = 50$. 

As expected from previous results, the majority of the perfect predictions for this task can be done only using T5 (\about 93\%). A limited percentage of perfect predictions is shared (\about 5\%), and a minority of instances can be only predicted through the baseline (\about 2\%). 

%
%

\begin{figure}
	\textbf{Automatic Bug Fixing (BF)}
	\vspace{-0.12cm}
	\begin{lstlisting}[escapechar=^]
public void method_1(int var_1, int var_2, Intent data) {
  super.method_1(var_1, var_2, data);
  if ((data != null) && ((data.method_2(var_3)) ^\textbf{!=}^ string_1)) {
    var_4.setText(data.method_2(var_3));
  }
}
	\end{lstlisting}
	\vspace{-0.4cm}
	\begin{lstlisting}[escapechar=^,style=mystyleresult]
^\textbf{>}^ if ((data != null) && (^\textbf{!}^(data.method_2(var_3)).^\textbf{equals}^(string_1))) {
	\end{lstlisting}
		
	\textbf{Injection of Code Mutants (MG)}
	\vspace{-0.12cm}
	\begin{lstlisting}[escapechar=^]
public void method_1(result) {
  if (result == null) {
    var_1.setEnabled(false);
    var_2.setEnabled(false);
    return;
  } 
  var_3.setText((^\textbf{string\_1}^ + result));
}
	\end{lstlisting}
	\vspace{-0.4cm}
	\begin{lstlisting}[escapechar=^,style=mystyleresult]
^\textbf{>}^ var_3.setText((^\textbf{string\_2}^ + result));
	\end{lstlisting}
		
	\textbf{Generation of Assert Statements in Test Methods (AG)}
	\vspace{-0.12cm}
	\begin{lstlisting}[escapechar=^]
//test method
void testSetPosition() {
  Position result = instant1.getPosition();
  Position p = org.geotools.temporal.Object.defaultPosition(new Date());
  ((org.geotools.temporal.Object.DefaultInstant)(instant1)).setPosition(p);
  ^\textbf{<assertplaceholder>}^;
}
//focal method
public Position getPosition() { return this; }
	\end{lstlisting}
	\vspace{-0.4cm}
	\begin{lstlisting}[escapechar=^,style=mystyleresult]
^\textbf{>}^ ^\textbf{assertFalse(instant1.getPosition().equals(result))}^
	\end{lstlisting}

	\textbf{Code Summarization (CS)}
	\vspace{-0.12cm}
	\begin{lstlisting}[escapechar=^]
public boolean forEachPair(final IntObjectProcedure procedure) {
  for (int i = table.length; i-- > 0;) {
    if (state[i] == FULL) 
      if (!procedure.apply(table[i], values[i]))
        return false;
  }
  return true;
}
	\end{lstlisting}
	\vspace{-0.4cm}
	\begin{lstlisting}[escapechar=^,style=mystyleresult]
^\textbf{>}^ ^\textbf{"applies a procedure to each key value pair of the receiver if any"}^
	\end{lstlisting}
	\caption{Examples of perfect predictions done by T5.}
	\label{fig:qualitative}
\end{figure}

\subsection{Qualitative Examples}

We show in \figref{fig:qualitative} four examples of perfect predictions by T5; for the sake of space limitations, for BF and MG, we only report the parts of code that T5 modified. In the first one (BF), the developers used the \texttt{!=} operator for comparing two objects instead of calling the \texttt{equals} method. T5 was able to fix the bug by (i) adding a call to \texttt{equals} and, less obvious, (ii) adding the \texttt{!} operator before the method call. 
In the second example (MG), T5 generates a mutant by replacing the correct string (\texttt{string\_1}) with a different one (\texttt{string\_2}).
In the third example (AG), the test checks if a call to \texttt{setPosition} on the variable \texttt{instant1} does not change its value (\ie it should be equal to \texttt{result}). Such an assertion is not trivial to generate since \texttt{result} is used in the assertion even if \texttt{p} is closer to the assert placeholder. 
Finally, in the last example (CS), T5 detects that (i) the method applies a given procedure to the pairs, (ii) the pairs belong to a receiver, and (iii) this happens only if there is a receiver. This shows how T5 is able to generate a summary for a method that even a developer could struggle to understand.


\subsection{Answer to our Research Question}
\label{sub:answer}
\revised{Our study showcases the potential of T5 for code-related tasks. The T5 model achieved better performance as compared to all baselines we experimented with. However, it is important to highlight that there are many factors that may have contributed to such a result. Indeed, the high effectiveness we obtained on all the tasks we experimented with might not only be related to the T5 architecture (\eg the fact that the T5 supports transfer learning with knowledge acquired on a task that can be reused on other tasks) but to other differences between the study presented in this paper and the experiments performed in the original work. While the datasets used for testing the techniques are exactly the same, two aspects must be considered. First, the type of the model we use (\ie the transformer model): using such a model in a single-task setting may still allow to achieve an improvement over the respective baselines 
Second, as previously explained, the pre-training phase may provide ``knowledge'' to the model that is not available in the training sets used for the fine-tuning (and, thus, not used by the competitive techniques).}

The results in terms of inference time show that T5 is able to complete all the tasks very quickly: it always takes less than 6 seconds even to generate 50 alternative solutions. 

Note that the inference times we reported are based on the usage of a consumer-level device and by only using CPUs: when using GPUs (Nvidia Tesla P100 provided by Google Colab), the time needed for each task flattens to at most \about 0.5 seconds for $K = 50$, \ie the task variability previously reported disappears. Finally, the overlap analysis indicates that some instances of the considered tasks were not resolved correctly by T5 but were resolved by the baselines. This means that such instances can be still resolved automatically by a ML-based approach. Such a consideration suggests that there is still room for improving the accuracy of T5. 

\section{Threats to Validity} \label{sec:threats}
\textbf{Construct validity.} For both the pre-training and the fine-tuning of our model we re-used available datasets, just performing some additional cleaning (\eg removal of duplicates after abstraction in the dataset used for the pre-training). 
Even though we remove duplicates from the pre-training dataset and  double-check all the datasets used for the fine-tuning, it is possible that instances in the pre-training dataset appear in some of the test datasets we reused. For example, a code comment included among the pre-training instances we used could have a duplicate, by chance, in the test set of the CS task, thus helping the T5 model in the prediction. While removing instances from the test sets was not an option since this would not allow a fair comparison between the T5 results and the ones reported in the original papers, we decided to investigate such overlap, to have an idea of the extent to which it could have influenced our findings. We found 0 duplicates between the pre-training dataset and the test sets of: \BFsmall, \AGabs, \AGraw; 1 in \MGident; 2 in \BFmedium; and 147 (out of 90,908) in the \CS test set. Thus, the influence of duplicates on the reported results should be marginal. 

\textbf{Internal validity.} An important factor that influences DL performance is hyperparameters tuning. For the pre-training phase, we used the default T5 parameters selected in the original paper \cite{raffel2019exploring} since we expect little margin of improvement for such a task-agnostic phase. For the fine-tuning, due to feasibility reasons, we did not change the model architecture (\eg number of layers) but we experiment with different learning rates. We are aware that a more extensive calibration would likely produce better results.

\textbf{External validity.} We experimented the T5 model on four tasks using six datasets. The main generalizability issue is related to the focus on Java code. However, excluding the abstraction component, our approach is language agnostic.

\section{Conclusion} \label{sec:conclusion}
\revised{We investigated the usage of a T5 model to support four code-related tasks: \textit{automatic bug-fixing}, \textit{generation of assert statements in test methods}, \textit{code summarization}, and \textit{injection of code mutants}. The achieved results  show that the T5 model can be successfully used for these tasks, with performance superior to the four  baselines. However, as explained in \secref{sub:answer}, such a finding deserves additional investigations TO better understand what makes T5 performing better.} 

Also, Raffel \etal \cite{raffel2019exploring}, who originally introduced T5, showed that larger T5 models are able to achieve much better results as compared to the \textit{small} T5 model we used in this work. From this perspective, the results reported in this paper should be considered as a lower bound of the T5 capabilities.

Code and data used in this paper are publicly available \cite{replication}.




\section*{Acknowledgment}
This project has received funding from the European Research Council (ERC) under the European Union's Horizon 2020 research and innovation programme (grant agreement No. 851720). W\&M team was supported in part by the NSF CCF-1955853 and CCF-2007246 grants. Any opinions, findings, and conclusions expressed herein are the authors' and do not necessarily reflect those of the sponsors. 

\balance

\bibliography{main}
\bibliographystyle{IEEEtran}

\end{document}